# Perceptions of the Fourth Industrial Revolution and AI's Impact on Society

Daniel A. Agbaji
danielagbaji@my.unt.edu

Brady D. Lund

And

Nishith Reddy Mannuru

University of North Texas, College of Information

Denton, TX, USA

**Abstract:**
The Fourth Industrial Revolution, particularly Artificial Intelligence (AI), has had a profound impact on society, raising concerns about its implications and ethical considerations. The emergence of text generative AI tools like ChatGPT has further intensified concerns regarding ethics, security, privacy, and copyright. This study aims to examine the perceptions of individuals in different information flow categorizations toward AI. The results reveal key themes in participant-supplied definitions of AI and the fourth industrial revolution, emphasizing the replication of human intelligence, machine learning, automation, and the integration of digital technologies. Participants expressed concerns about job replacement, privacy invasion, and inaccurate information provided by AI. However, they also recognized the benefits of AI, such as solving complex problems and increasing convenience. Views on government involvement in shaping the fourth industrial revolution varied, with some advocating for strict regulations and others favoring support and development. The anticipated changes brought by the fourth industrial revolution include automation, potential job impacts, increased social disconnect, and reliance on technology. Understanding these perceptions is crucial for effectively managing the challenges and opportunities associated with AI in the evolving digital landscape.

# Perceptions of the Fourth Industrial Revolution and AI's Impact on Society

**Abstract:**
The Fourth Industrial Revolution, particularly Artificial Intelligence (AI), has had a profound impact on society, raising concerns about its implications and ethical considerations. The emergence of text generative AI tools like ChatGPT has further intensified concerns regarding ethics, security, privacy, and copyright. This study aims to examine the perceptions of individuals in different information flow categorizations toward AI. The results reveal key themes in participant-supplied definitions of AI and the fourth industrial revolution, emphasizing the replication of human intelligence, machine learning, automation, and the integration of digital technologies. Participants expressed concerns about job replacement, privacy invasion, and inaccurate information provided by AI. However, they also recognized the benefits of AI, such as solving complex problems and increasing convenience. Views on government involvement in shaping the fourth industrial revolution varied, with some advocating for strict regulations and others favoring support and development. The anticipated changes brought by the fourth industrial revolution include automation, potential job impacts, increased social disconnect, and reliance on technology. Understanding these perceptions is crucial for effectively managing the challenges and opportunities associated with AI in the evolving digital landscape.

## Introduction

The Fourth Industrial Revolution, specifically driven by Artificial Intelligence (AI), has had a profound and wide-ranging impact on various aspects of our lives (Spöttl & Windelband, 2021; Hyun Park et al., 2017). This significant influence has prompted individuals from diverse backgrounds and demographics to ponder which social, economic, and political factors will play the most substantial role in shaping the use of AI and whether its overall impact will be beneficial or detrimental (Kaplan & Haenlein, 2020; Petit, 2017; Prince & Schwarcz, 2019). Adding to these concerns, the recent advancement of Text Generative AI tools like ChatGPT has raised significant ethical, security, privacy, and copyright issues, as the content produced by such tools raises questions and challenges for users (West, 2018; Coeckelbergh, 2020; Vartiainen & Tedre, 2023).

While previous research has explored the influence of user perceptions on various aspects of society, including social dynamics, politics, the economy, and education, most studies have focused on understanding why and not how users form their perceptions of AI (Williams et al., 2019; Wood & Evans, 2018). To address these gaps, this study examines individuals in different information flow categorizations, specifically exploring how their perceptions of AI may differ and how these perceptions can impact their interactions with AI technology.

To help bridge the gaps in current studies relating to the perception of AI among users, this study used data from semi-structured interviews to analyze the current perceptions of AI among varying populations. The results showed what level of information flow an AI user will likely fall under, their sources of information relating to AI, and how their perception of AI affects its engagement and recommendation. It also revealed concerns AI users have and their suggestions to solve some of the current social-economical, privacy, and copyright issues relating to the use of AI.

## Literature Review

The term "Industrial Revolution" refers to a period of rapid industrial and economic growth and societal change. We are on the verge of a technology revolution, that will radically transform the way we live, work, and interact with one another, while also bringing major changes in the world around us.

The First Industrial Revolution was characterized by the transition from manual labor to mechanized production. It changed agrarian societies to industrialized ones as machines were made with the help of water and steam power (Schwab, 2016). The Second Industrial Revolution experienced fast industrial expansion that was much more advanced than the First Industrial Revolution resulting in significant improvements across a wide range of industries, including manufacturing, agriculture, and transportation (Fomunyam, 2019). Additionally, The second one made mass production possible with the help of electricity (Schwab, 2016). Hence this revolution was also called the Technological Revolution. The Third Industrial Revolution also referred to as the digital revolution, was characterized by the shift away from mechanical and analog technologies to digital electronics (Sakhapov & Absalyamova, 2018). That is, the Third used computers and gadgets to make the output more efficient (Schwab, 2016). Now, the Fourth

Industrial Revolution is constructed on the Third, which has been going on since the middle of the last century and is all about digital technology. The term Fourth industrial revolution also called Industry 4.0 or 4IR was coined by Klaus Schwab, who was the founder and the executive chairman of the World Economic Forum, and described it as "a world where individuals move between digital domains and offline reality with the use of connected technology to enable and manage their lives" (Xu et al., 2018,p.90). It is made up of a mix of technologies that blur the lines between the real, digital, and living worlds (Schwab, 2016). The use of quantum computing, artificial intelligence, machine learning, the Internet of Things (IoT), biotechnology, 3D printing, and autonomous vehicles are among the essential components. Even though each industrial revolution is often seen as an independent event, but they are better understood as a set of events that build on the ideas of the previous revolution where each revolution builds upon the inventions of the one before it and lead to more advanced innovations.

Throughout history, each industrial revolution brought about significant social changes. These innovations, which ranged from the steam engine to the Internet to AI, had a significant impact on the way the economy, jobs, and social life. And, only if we know about these changes and how fast they are happening, their advantages and disadvantages, can we make sure that improvements in knowledge and technology reach everyone and help everyone. Morrar et al., (2017) laid down a few reasons why Industry 4.0 is significant and is seen as revolutionary in this day and age of information technology. Firstly, Industry 4.0 makes it easier for manufacturers to deal with problems or challenges by making companies more flexible and able to adapt to business trends. Secondly, Industry 4.0 makes it possible for economies to undergo a change that will make them more innovative and, as a result, enhance their level of production. Third, it emphasizes the consumer's role as a co-producer and puts them at the center of everything. Finally, the application of cutting-edge technology to problems in areas like energy, resources, the environment, and social and economic repercussions will, in the end, pave the way for long-term success.

Recent literature on the topic of the Fourth Industrial Revolution has mostly concentrated on the technical revolutionary aspects of Industry 4.0. The fast pace of technological progression and digitalization has raised questions about its impact on people personally and on the whole of society. Furthermore, Brynjolfsson and McAfee (2014) stressed the significance of understanding how the new industrial revolution affects the whole society in order to capitalize on the opportunities it offers. Hence, it is necessary to view technology innovations from a social standpoint when thinking about how they may be used to address societal issues and also the impact and influence it has on one's life.

**IMPACT FROM DIFFERENT SOCIETAL STANDPOINTS:**

There is a widespread fear that new technologies will cause the loss of many jobs in the near future. This fear stems from the fact that more human duties can now be handled by machines. Bessen (2019) estimates that anywhere from 9% of jobs to 47% are at risk of automation in the near future. However, Arntz et al., (2016) suggest that AI will also create new job types and businesses, just like other economic changes and industrial revolutions did. These new jobs may need different skill sets and, if neglected, may widen socioeconomic gaps. Additionally, AI's ethical implications are also of major significance. As AI becomes more important in everyday life, problems like privacy, data security, bias, decision-making openness, and responsibility have become very important (Cath, 2018). Concerns have also been raised about the use of AI in surveillance, disinformation, and warfare. It is essential to address these ethical issues in order to prevent possible damage and to preserve the confidence of society. The effects of AI on education and learning are also very important to look at. AI will positively change education through personalized learning, more efficient administration, and global connections (Wang et al., 2023). However, there are worries about the privacy of data, the digital divide, and the need for educational institutions to adapt to fast-changing job markets. Different stakeholders, including teachers, students, and parents, have expressed opposing emotions (Luckin, 2018). Furthermore, understanding how governments react to the rise of advanced AI technologies is also of major importance. The pace of AI advancement and implementation may be significantly influenced by policy and regulatory responses (Schwab, 2016). Chen, (2009) calls for bold steps to guide the development of AI in a responsible way i.e., finding a balance between encouraging innovation and protecting society from its possible bad effects.

Overall, the 4IR, driven by AI and other related technologies, presents both opportunities and challenges. Understanding the diverse societal perceptions and potential impacts can help navigate the transformative changes these technologies bring. Therefore, in order to understand the perspectives of people with regard to the 4IR and the advancements in AI, a qualitative analysis has been conducted to better understand their views from different lenses.

## Methods

This study implemented a qualitative research method involving a semi-structured interview. Participants were contacted via email recruiting messages following an IRB approval of the study. A total of 15 participants were interviewed, during which 11 questions were asked to get relevant feedback on their information flow and perceptions of AI. The Zoom online video conferencing platform was used to conduct the interviews as it allows for both recording of the interview and transcription of the responses.

At the end of the interview process, participants' responses were coded to formulate relevant themes matching the purpose of the study. These themes were saved on an Excel spreadsheet allowing for easy analysis. Excel data analysis features were used to analyze the coded themes and for creating the visualization.

## Results

The results section of this paper is divided into two parts. In the first part, we look at themes in the participant-supplied definitions of the terms *artificial intelligence* and the *fourth industrial revolution*. In the second part, we look at responses to questions related to perceptions about AI and the fourth industrial revolution.

### Definitions of Key Terms

The following definitions were supplied by the participants for the terms artificial intelligence and fourth industrial revolution. Using the definition of AI provided by IBM (2023), as

"Artificial intelligence leverages computers and machines to mimic the problem-solving and decision-making capabilities of the human mind," and definition of the fourth industrial revolution by Schwab (2023) as "the convergence of digital, biological, and physical innovations," we calculated cosine similarity scores for each of the participant-supplied definitions. These similarity scores are provided in parentheses at the end of each definition, where a score of 1 indicates an identical meaning and a score of 0 indicates a completely dissimilar meaning.

*Definitions of Artificial Intelligence:*

1. It's a branch of computer science that deals with developing systems or machines with human-like intelligence. It includes areas like natural language processing. (.884)
2. It's like a combination of different technologies that have been developed to match human intelligence. so that they could even sense and reason, like a typical human, would do in any situation or instance. (.911)
3. Training a machine or giving a machine the authority of sense that to make a decision like humans. (.713)
4. I wouldn't say replicating human intelligence, but I would say it's like an offshoot of that where you're trying to replicate. I think it's more like automation is what I think. But instead of having like these big industrial machines, you're doing it using a computer. And previously you had these machines doing physical tasks, but these are more mental. (.530)
5. I would say that that's something that these are machines and computers doing tasks that humans would. (.616)
6. Any machine that can learn and solve problems like a human brain can. (972)
7. Mimicking human intelligence by computer. (.866)
8. I would simply define it as programming machines to replicate human intelligence. (.883)
9. It is something that has made lives a lot easier for people and take decisions on behalf of humans. (.614)
10. It's a system that tries to mimic human intelligence- to think as humans. (.866)
11. It's a digital computer where it performs all the tasks, which a normal human can perform. (.669)
12. The generalized definition would be like the stimulation of human intelligence into something artificial like robots or into a machine. I would like to phrase it in a different way where we try to teach something which doesn't have knowledge of it at all. (.750)
13. Artificial intelligence is that it's usually computer-based intelligence, computers have been taught to appear to speak and communicate and listen and understand like humans. (.956)
14. It's the simulation of human intelligence and machines and using machine learning algorithms that aim to mimic human cognitive thinking and to perform specified tasks. (.946)
15. I would define artificial intelligence as any non-human non-biological intelligence. (.365)

*Definitions of Fourth Industrial Revolution:*

1. The evolution of emerging technologies and computing. (.684)
2. I believe that it's about the digitalization and automation of the processes which even more replace the human activities. (.708)
3. Not Sure.
4. Using AI robotics and the information communication technologies to its major extent to improve the human life. (.868)
5. I think it's this whole ideal, or I don't want to make it sound like where humans are getting replaced. But it's kind of like computers are. I don't want to use the word replace, but they are trying to kind of invade our space. (.373)
6. I would say that it's more of like the automation and AI, the Internet of things kind of just this whole, maybe even like the social aspect of technology. (.692)
7. The integration of digital technologies into the manufacturing industry. (.759)
8. An era associated with increased technological innovations that eliminate physical, digital and biological restrictions which changes the way we work, live and relate with one another. (.876)
9. Don't Know.
10. It's about deep learning, an offset of AI built upon big data. (.668)
11. It's basically the new technological developments that are coming in this industry, such as virtual reality, artificial intelligence, and big data, which are growing currently. (.798)
12. Information being used, handled, and enhancing the outcome. (.450)
13. I kind of have an idea that it's related to using artificial intelligence. (.593)
14. I'd say that it is technological advancement. And it's an extension of the third revolution maybe, with more sophisticated programs, and all like, robotics, 3d printing, face recognition, and it's developing really quickly. And artificial intelligence-based systems and all that. (.911)
15. Not Sure.

## Synthesized Definitions

With the assistance of ChatGPT, the following definitions were derived based on themes in the participants individual definitions of artificial intelligence and the fourth industrial revolution:

*Artificial intelligence* is the development and use of digital/computer-based intelligence to replicate human abilities, through machine learning and automation, with the goal of enhancing human life by automating tasks and expanding intelligence beyond human capabilities.

The *Fourth Industrial Revolution* is a transformative period marked by the integration of digital technologies into industries, revolutionizing processes through digitalization, automation, and emerging technologies, impacting human activities and enhancing information handling and outcomes.

The following subsections identify some key themes that emerge in the participant-supplied definitions for each term.

## Themes in Definitions of Artificial Intelligence

1. Replicating human intelligence: Several responses mention the idea of artificial intelligence aiming to replicate or mimic human intelligence. This includes aspects such as sensing, reasoning, learning, problem-solving, and decision-making.

2. Machine learning and automation: The concept of training machines, programming them, or using algorithms to enable learning and problem-solving is a common theme. Many responses highlight the role of automation, replacing physical tasks with mental tasks performed by machines.

3. Human-like tasks and capabilities: Artificial intelligence is often associated with machines performing tasks that humans can do, such as speaking, communicating, listening, understanding, and solving problems. The focus is on achieving human-like abilities through technology.

4. Digital/computer-based intelligence: The idea that artificial intelligence exists in the realm of digital computers is mentioned in a couple of responses. It emphasizes that AI systems are built using computer technology.

5. Enhancement and ease of life: Some responses highlight the positive impact of artificial intelligence on human lives, making tasks easier and enabling machines to make decisions on behalf of humans. This theme emphasizes the benefits of AI technology.

6. Non-human, non-biological intelligence: One response offers a broader definition of artificial intelligence as any intelligence that is not human or biological in nature. This theme touches on the concept of intelligence existing beyond human capabilities.

*Themes in Definitions of Fourth Industrial Revolution*

1. Digitalization and Automation: The notion of digitalizing and automating processes, often to replace human activities, is a recurring theme. It highlights the role of emerging technologies and computing in transforming traditional systems.

2. Technological Advancements: Several responses mention the Fourth Industrial Revolution as a period characterized by technological developments. This includes AI, robotics, virtual reality, big data, deep learning, the Internet of Things, and sophisticated programs.

3. Integration of Digital Technologies: The integration of digital technologies into various industries, such as manufacturing, is highlighted as a key aspect of the Fourth Industrial Revolution.

4. Changing the Way We Work and Live: The idea that the Fourth Industrial Revolution brings about changes in the way we work, live, and relate to one another is mentioned. It signifies the transformative impact of technological innovations on society.

5. Potential Replacement of Humans: Some responses express concerns or allude to the idea that computers or technology might replace or invade human spaces. This theme touches on the evolving relationship between humans and technology.

6. Enhanced Information Handling and Outcomes: The use of information, handling of data, and the potential for enhancing outcomes through technology are mentioned as relevant aspects of the Fourth Industrial Revolution.

**Perceptions of Artificial Intelligence and the Fourth Industrial Revolution**

*Extent Informed about Artificial Intelligence*

Figure 1 displays the responses to the extent to which participants felt informed about trends in artificial intelligence. The majority of respondents indicated that they were at least somewhat well informed about these trends; however, a few of the respondents indicated that they were not very well informed or were only informed to the extent to which they received information from a single platform, such as Facebook or Tik Tok.

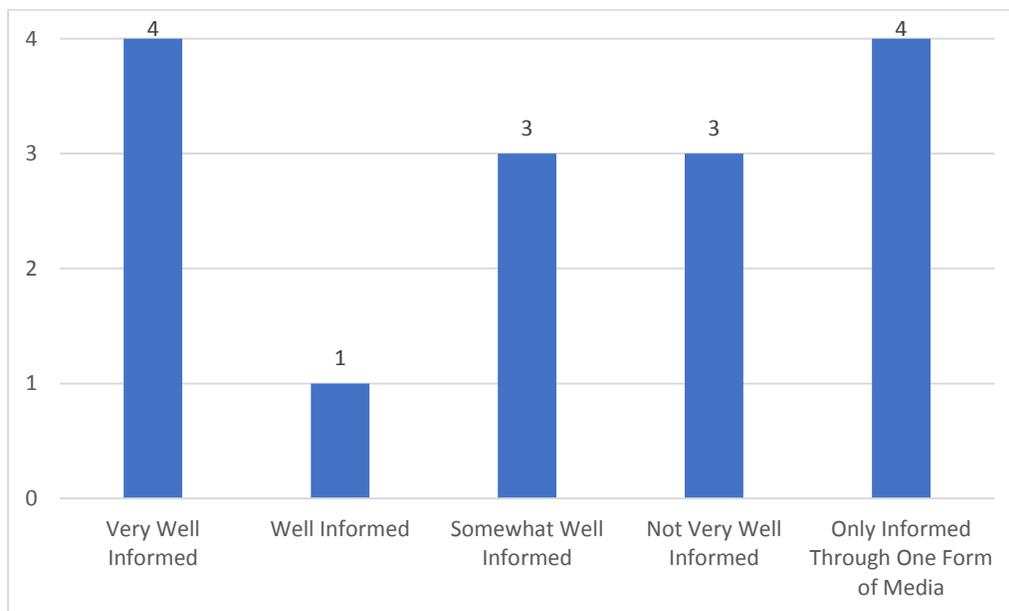

*Where Information About Use of AI Received*

Figure 2 shows the results for where information about AI was received by participants. News media and social media were clearly the most common sources. Sources like scholarly journals and blogs were mentioned by a few participants, particularly those who indicated that they were "very well informed" about artificial intelligence. While these results are presented in aggregate categories, such as "news media" and "social media," most respondents named more-specific sources. For instance, for "news media," three participants named "news websites," two named "newspapers," three named "television news," and one named "public radio." For social media, one named "Instagram," two named "LinkedIn," one named "Twitter," and four listed "Social Media Posts (general)."

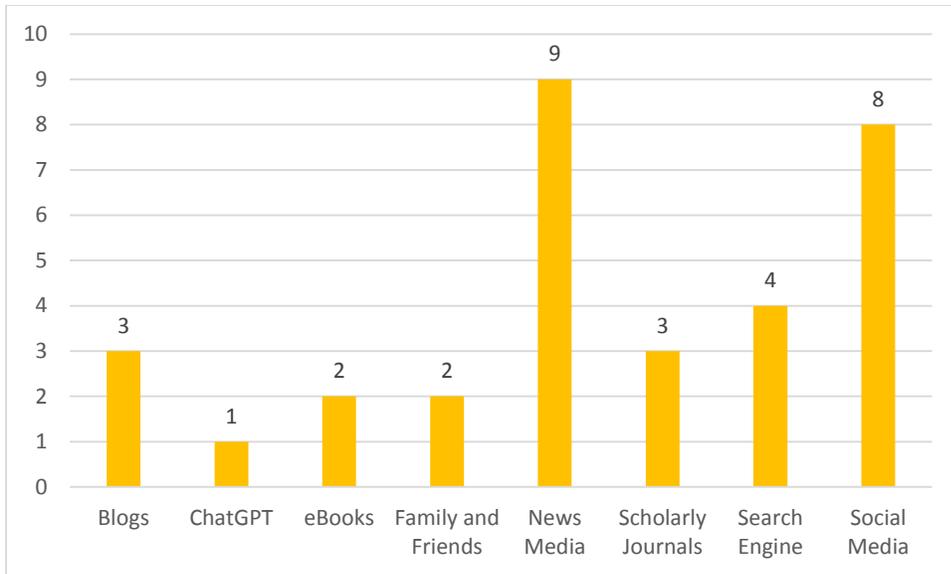

*Belief Receiving Accurate AI Information*

Belief about receiving accurate AI information was nearly evenly split between those who did believe the information was accurate and those who believed that the information was sometimes accurate and sometimes inaccurate. Interestingly, those who felt they were very well informed or not very well informed tended to feel that they sometimes received accurate information, while those who felt they were only somewhat well informed or only informed through one form or media were more confident that they always received accurate information. This may be representative of 1) a lack of awareness about misinformation related to AI among those who are very well informed, 2) a (potentially warranted) lack of confidence about AI informed among those who are not very well informed, and 3) a potential overestimation of the reliability of AI information among those who feel only somewhat informed.

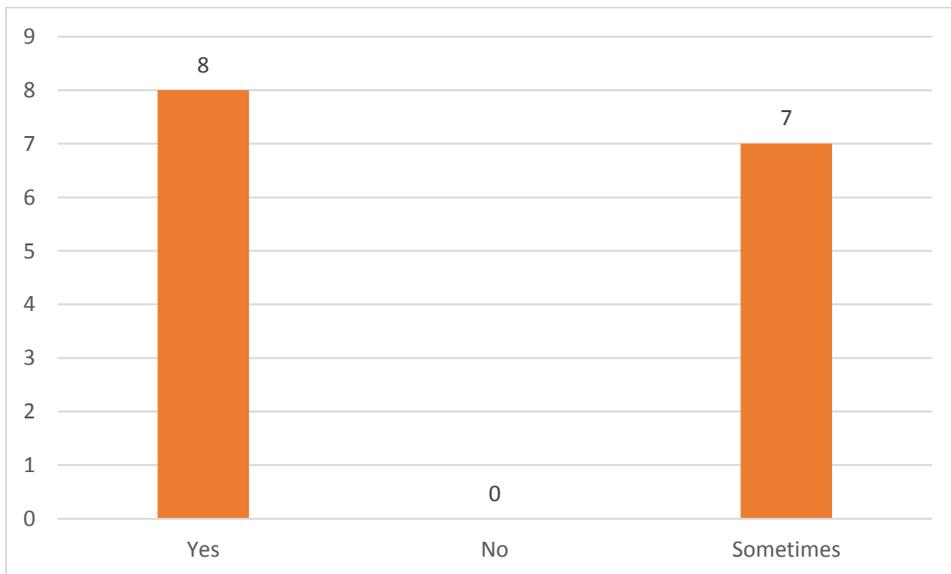

*Concerns about AI Technologies*

Figure 4 displays the responses for participants concerns about AI technologies. The two greatest concerns were job replacement – the belief that AI may eliminate human jobs – and invasions of privacy against those who interact with AI interfaces. The inaccuracy of information provided by AI was also a common concern. This issue has been regularly reported with current generative AI technology such as ChatGPT (Teel, Wang, & Lund, 2023). Additional concerns focus on elements of deception or bias contained within AI algorithms and the potential for AI replacing human ingenuity. Those who felt less informed about AI were particularly concerned about the issue of job replacement, while those who felt very well informed about AI were particularly concerned about invasions of privacy, data bias, and inaccurate information.

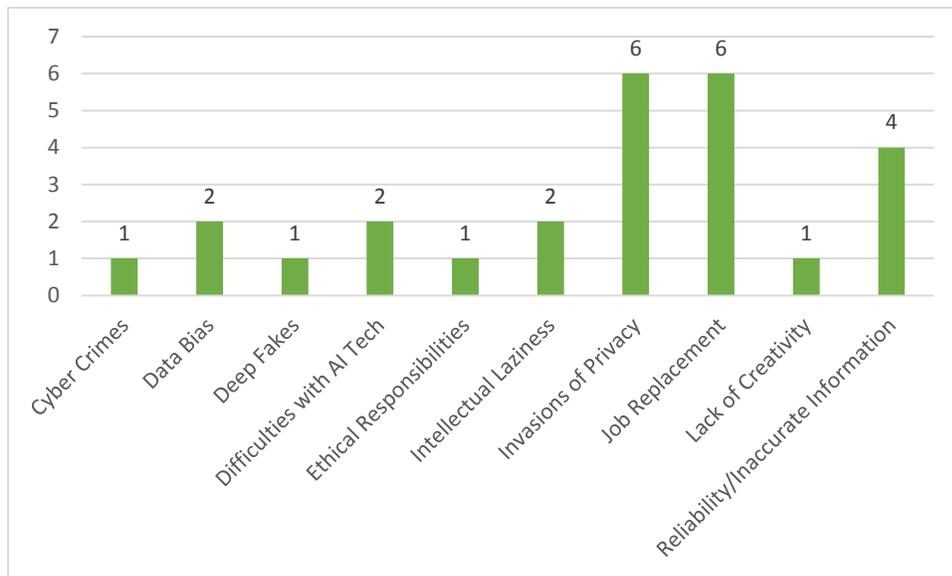

*Potential Benefits and Drawbacks of the 4IR*

Shown in Figure 5 are the potential benefits and drawbacks of the fourth industrial revolution noted by participants. The majority of the items named by participants were benefits (68% of items were benefits compared to 32% drawbacks). The ability to make human lives easier through solving complex problems and increasing convenience were the most commonly named benefits. Some respondents also noted 4IR technology could work along with humans in order to increase the speed and efficiency of their efforts. As with AI technology in general, among the biggest concerns with the 4IR were privacy and bias issues, along with concerns about how some 4IR technologies may result in reduced connectedness among people as they rely on mobile technology and AI interfaces.

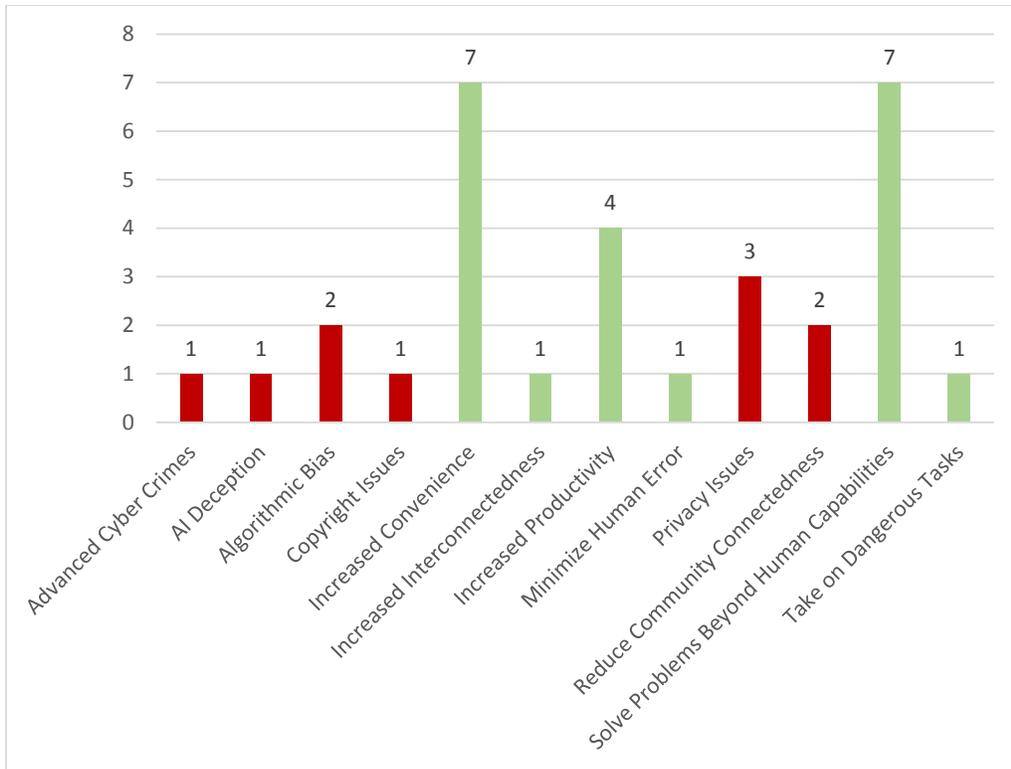

### Government Involvement in Shaping 4IR

Participants were deeply divided on how the government should be involved in shaping the fourth industrial revolution, as shown in Figure 6. While fives participants indicated that strict regulations should be placed on 4IR innovators at the federal or state/local level, over half of the participants indicates that little or no regulations should be placed and that, instead, governments should encourage 4IR development through funding, outreach, and public-private partnerships. A few respondents indicated that both strict regulation and support and development should be encouraged.

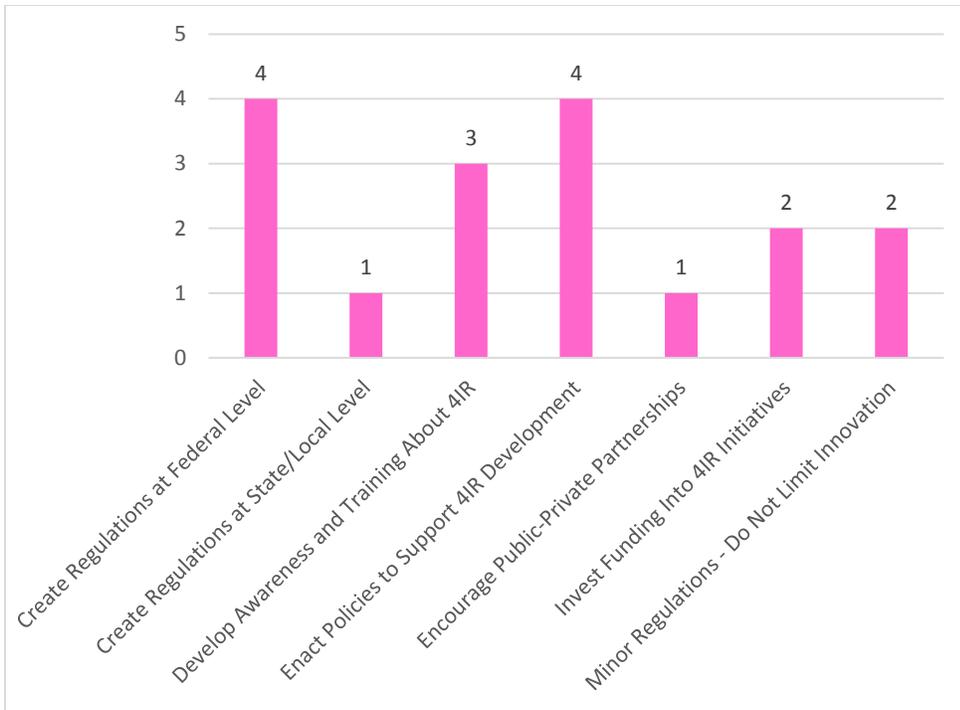

## How 4IR Will Change Ways We Live and Work

Finally, Figure 7 illustrates the participant responses as to how the 4IR will change the ways that people live and work. While many participants indicated that greater automation would be an inevitable outcome, most did not necessarily equate this development to a decrease in well-paying jobs. Several participants noted likely outcomes related to greater social disconnect among humans and greater reliance and addiction to technology.

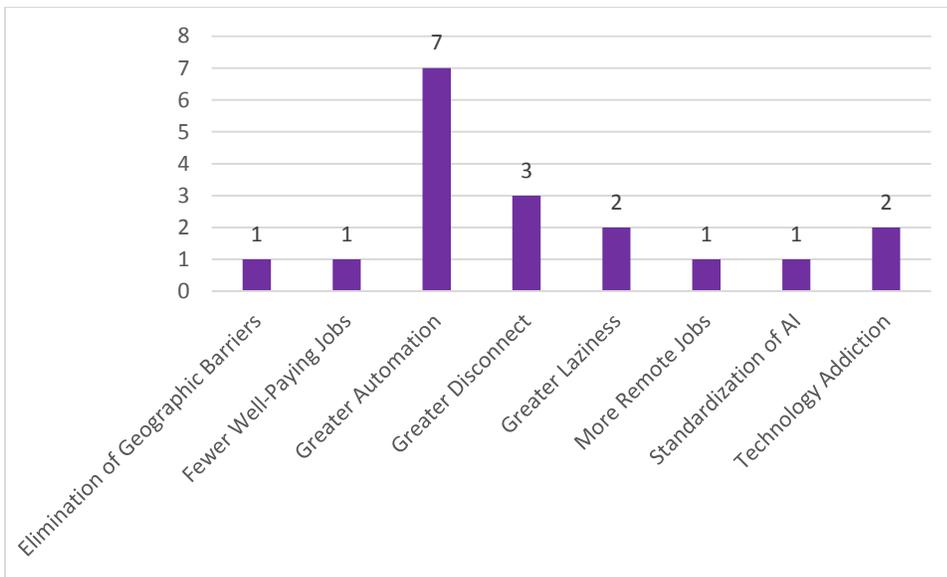

## Discussion

The analysis of participant-supplied definitions revealed several key themes related to artificial intelligence. The majority of responses emphasized the goal of replicating human intelligence through machine learning and automation. Participants highlighted the ability of AI to perform human-like tasks and enhance human life by automating tasks and expanding intelligence beyond human capabilities. The digital/computer-based nature of AI and its positive impact on making lives easier were also commonly mentioned themes. Some participants offered broader definitions of AI, perceiving it as non-human, non-biological intelligence. As is evident from these responses, AI can be defined and perceived in many ways depending on the individual, their background and knowledge of the technology. There may not be a single definition that is collectively understood by all people, and, indeed, this has been reflected in past studies as well (Helm et al., 2020). Nonetheless, there are commonalities in how AI in perceived as a technology that augments human skills or capacities (Legg & Hutter, 2007).

Regarding the definitions of the fourth industrial revolution, participants focused on various aspects of this emerging revolution. They frequently mentioned the digitalization and automation of processes, the integration of digital technologies into industries, and the transformative impact on work, life, and relationships. Technological advancements such as artificial intelligence, robotics, virtual reality, big data, and deep learning were also frequently associated with the fourth industrial revolution. Concerns were raised about potential job replacement, invasions of privacy, and the accuracy of information provided by AI systems. Fears about 4IR technologies like AI are not uncommon and have been noted in prior research focusing on the perceptions of lay people toward the technology (Li & Huang, 2020; Sindermann et al., 2022). However, participants also recognized the potential benefits of the 4IR, such as solving complex problems and increasing convenience, which align with benefits identified in recent literature (Wang et al., 2023).

Participants' perceptions of artificial intelligence and the fourth industrial revolution were influenced by their level of knowledge and the sources of information they relied upon. Most participants considered themselves at least somewhat well informed about AI trends, with news media and social media being the primary sources of information. However, concerns were raised about the accuracy of the information received, with participants indicating varying degrees of belief in its accuracy. Participants who felt more informed expressed concerns about privacy, data bias, and inaccurate information, while those who felt less informed were particularly concerned about job replacement. These findings have been substantiated in past studies to some extent (Lund et al., 2020), but this study provides a true comparison of both respondents who are innovators and those who are reluctant adopters.

Opinions on government involvement in shaping the fourth industrial revolution were divided among participants. While some favored strict regulations at the federal or state/local level, a majority believed that governments should encourage 4IR development through funding, outreach, and public-private partnerships. A few participants suggested a balanced approach, advocating for both strict regulation and support for development. The role of government in the fourth industrial revolution has been a contested issue in the literature as well as industry (Shava

& Hofisi, 2017; Xu et al., 2018). While the findings of this study do not provide solutions, they do offer some indication of the perspectives of a diverse population of individuals.

Finally, participants anticipated that the fourth industrial revolution would lead to greater automation and changes in the ways people live and work. While some participants expressed concerns about job loss, others did not necessarily associate automation with a decrease in well-paying jobs. Participants also mentioned potential social disconnect and increased reliance on technology as likely outcomes. The risk of automation impacting jobs and communication are common concerns among individuals in a variety of industries (Lund, 2021). However, many posit that these risks will also come with opportunity for growth in new professions.

This study has several limitations that should be acknowledged. Firstly, the participant-supplied definitions of artificial intelligence and the fourth industrial revolution were based on a specific sample of individuals, and their perspectives may not be representative of the broader population. Additionally, the cosine similarity scores used to evaluate the definitions may not fully capture the nuances and variations in meaning. Furthermore, the study's focus on examining individuals in different information flow categorizations may limit the generalizability of the findings. The reliance on self-reported levels of information and accuracy perception regarding AI trends introduces potential biases and subjective interpretations. Finally, the survey responses may be influenced by social desirability bias, as participants may provide responses, they believe are more socially acceptable or align with popular opinions.

## Conclusion

This study found that participants' perceptions of AI and the fourth industrial revolution are influenced by their sources of information and level of knowledge concerning AI. The findings of this study suggest that the most visited sources of AI information are the News and social media platforms, and that the more information individuals have on AI, the more likely they are receptive to the use of AI and other tools provided by the fourth industrial revolution. In contrast, low information about AI means low likelihood of accepting the use of AI.

Consequently, this study reveals that job loss is the main concern participants have concerning using AI and other tools provided by the 4IR. However, participants highlighted more benefits in the use of AI than the concerns they have shared. With regards to the future trajectory of AI in social-economic activities, the study reveals that more automation and digitization of activities that usually involve humans are likely to be replaced by AI. Further study using a qualitative research approach is likely to provide sufficient data that can be used to generalize the findings of this study.

**References**:

Arntz, M., Gregory, T., & Zierahn, U. (2016). The risk of automation for jobs in OECD countries: A comparative analysis.

Bessen, J. (2019). Automation and jobs: When technology boosts employment. *Economic Policy*, *34*(100), 589-626.

Brynjolfsson, E., & McAfee, A. (2014). *The second machine age: Work, progress, and prosperity in a time of brilliant technologies*. WW Norton & Company.

Cath, C. (2018). Governing artificial intelligence: Ethical, legal and technical opportunities and challenges. *Philosophical Transactions of the Royal Society A: Mathematical, Physical and Engineering Sciences*, *376*(2133), 20180080. https://doi.org/10.1098/rsta.2018.0080

Chen, H. (2009). AI, E-government, and Politics 2.0. *IEEE Intelligent Systems*, *24*(5), 64–86. https://doi.org/10.1109/MIS.2009.91

Coeckelbergh, M. (2020). AI ethics. Mit Press.

Fomunyam, K. G. (2019). Education and the Fourth Industrial Revolution: Challenges and possibilities for engineering education. *International Journal of Mechanical Engineering and Technology*, *10*(8), 271–284.

Helm, J. M., Swiergosz, A. M., Haeberle, H. S., Karnuta, J. M., Schaffer, J. L., Krebs, V. E., Spitzer, A. I., & Ramkumar, P. N. (2020). Machine learning and artificial intelligence: Definitions, applications, and future directions. *Current Reviews in Musculoskeletal Medicine, 13*, 69-76.

IBM. (2023). What is artificial intelligence (AI)? Retrieved from https://www.ibm.com/topics/artificialintelligence#:~:text=What%20is%20artificial%20intelligence%20(AI)%3F,capabilities%20of%20the%20human%20mind

Kaplan, A., & Haenlein, M. (2020). Rulers of the world, unite! The challenges and opportunities of artificial intelligence. Business Horizons, 63(1), 37-50.

Legg, S., & Hutter, M. (2007). A collection of definitions of intelligence. Proceedings of the Conference

on Advances in Artificial General Intelligence, 2006, 17-24.

Li, J., & Huang, J. (2020). Dimensions of artificial intelligence anxiety based on the integrated fear acquisition theory. *Technology in Society, 63*, article 101410.

Luckin, R. (2018). *Machine Learning and Human Intelligence: The future of education for the 21st century*. UCL IOE Press. UCL Institute of Education, University of London, 20 Bedford Way, London WC1H 0AL.

Lund, B. D. (2021). The fourth industrial revolution. *Information Technology and Libraries, 40*(1), 1-4.

Lund, B. D., Omame, I., Tijani, S., & Agbaji, D. (2020). Perceptions toward artificial intelligence among academic library employees and alignment with the diffusion of innovations' adopter categories. *College and Research Libraries, 81*(5), 865-882.

Morrar, R., Arman, H., & Mousa, S. (2017). The Fourth Industrial Revolution (Industry 4.0): A Social Innovation Perspective. *Technology Innovation Management Review*, *7*(11), 12–20.

Petit, N. (2017). Law and regulation of artificial intelligence and robots-conceptual framework and normative implications. Available at SSRN 2931339.

Prince, A. E., & Schwarcz, D. (2019). Proxy discrimination in the age of artificial intelligence and big data. Iowa L. Rev., 105, 1257.

Sakhapov, R., & Absalyamova, S. (2018). Fourth industrial revolution and the paradigm change in engineering education. *MATEC Web of Conferences*, *245*, 12003. https://doi.org/10.1051/matecconf/201824512003

Schwab, K. (2016, January 14). *The Fourth Industrial Revolution: What it means and how to respond*. World Economic Forum. https://www.weforum.org/agenda/2016/01/the-fourth-industrial-revolution-what-it-means-and-how-to-respond/

Schwab, K. (2023). The fourth industrial revolution. Retrieved from https://www.britannica.com/topic/The-Fourth-Industrial-Revolution-2119734

Shava, E., & Hofisi, C. (2017). Challenges and opportunities for public administration in the fourth industrial revolution. African Journal of Public Affairs, 9(9), 203-215.

Sindermann, C., Yang, H., Elhai, J. D., Yang, S., Quan, L., Li, M., & Montag, C. (2022). Acceptance and fear of artificial intelligence: Associations with personality in a German and a Chinese sample. Discovery Psychology, 2, article 8.

Teel, Z. A., Wang, T., & Lund, B. D. (2023). ChatGPT conundrums: Probing plagiarism and parroting problems in higher education practices. *College and Research Libraries News, 84*(6), 205-207.

Vartiainen, H., & Tedre, M. (2023). Using artificial intelligence in craft education: crafting with text-to-image generative models. Digital Creativity, 34(1), 1-21.

Wang, T., Lund, B. D., Marengo, A., Pagano, A., Mannuru, N. R., Teel, Z. A., & Pange, J. (2023). Exploring the potential impact of artificial intelligence (AI) on international students in higher education: Generative AI, chatbots, analytics, and international student success. Applied Sciences, 13(11), article 6716.

West, D. M. (2018). The future of work: Robots, AI, and automation. Brookings Institution Press.

Williams, R., Park, H. W., & Breazeal, C. (2019, May). A is for artificial intelligence: the impact of artificial intelligence activities on young children's perceptions of robots. In Proceedings of the 2019 CHI Conference on Human Factors in Computing Systems (pp.1-11).

Wood, B. A., & Evans, D. (2018). Librarians' perceptions of artificial intelligence and its potential impact on the profession. Computers in Libraries, 38(1).

Xu, M., David, J. M., & Kim, S. H. (2018). The fourth industrial revolution: Opportunities and challenges. *International Journal of Financial Research, 9*(2), 90-95.